\documentclass{article}
\usepackage{epsfig}

\date{\today}

\usepackage{amsmath}

\newcommand{\la}{\lambda}
\newcommand{\ka}{\kappa}
\newcommand{\al}{\alpha}

\newcommand{\f}{\phi}

\newcommand{\ee}{\end{equation}}
\newcommand{\eea}{\end{eqnarray}}
\newcommand{\be}{\begin{equation}}
\newcommand{\bea}{\begin{eqnarray}}

\newcommand{\pa}{\partial}

\newcommand{\vep}{\varepsilon}

\begin{document}

\begin{center}
{\Large\bf ON ROTATING SOLITONS IN GENERAL RELATIVITY }
\vspace{1.cm}\\
Eugen Radu 
\\ \emph{Department of  Mathematical Physics,
National University of Ireland Maynooth, Ireland}
\end{center}
\begin{abstract}
We review a number of results
recently obtained in the area of constructing 
rotating solitons in a four dimensional asymptotically flat 
spacetime.
Various models are examined, special attention being paid to the 
monopole-antimonopole  and gauged skyrmion configurations, 
which have a nonvanishing total angular momentum.
For all known examples of rotating solitons,
the angular momentum is fixed by some conserved charge of the matter fields.
\end{abstract}

\section{Introduction}

Rotation is a universal
phenomenon, which seems to be shared by all objects, at all possible
scales. 
However, as shown first by Lichnerowicz \cite{Lichner},
the vacuum Einstein equations admit  no 
particle-like solutions, in particular no rotating regular 
configurations. 
For a gravitating Maxwell field, the Kerr-Newman black hole
solutions represent the only physically reasonable, asymptotically flat configurations with
nonzero angular momentum. However, again no regular rotating solution is
found in the limit of zero event horizon radius.

The inclusion of more general (nonlinear-) matter sources in the theory leads to
the possibility of finding localized, globally regular, particle-like objects 
with finite energy -- so-called solitons. 
Most of these solutions correspond to gravitating generalizations 
of the flat spacetime solitons, $e.g.$
 the monopoles \cite{mon},
 dyons \cite{dyon}, sphalerons \cite{sphal} and Skyrmions \cite{Skyrme:1961vq}.
There are also examples of solitons which have no Minskowski space counterparts, the 
Bartnik-McKinnon (BK) solutions \cite{Bartnik:1988am}
in Einstein-Yang-Mills (EYM) theory and the boson stars 
 \cite{Kaup, Ruffini:1969qy}   being the best known cases.

However, most of particle-like 
solutions discussed in the literature  
are spherically symmetric. Then it is natural to wonder whether one can 
find axially symmetric solitons with a nonzero angular momentum.
 This problem enjoyed recently some interest, the issue of rotating soliton 
solutions being systematically considered  for various models,
with some surprising results.
In this work we review the situation for several different
cases, presenting the basic features of  rotating solutions.
For simplicity, we'll restrict to asymptotically flat solutions
and four spacetime dimensions \footnote{ Here we do not ask the solitons to be stable; also
the spacetime is supposed to possess a  $R^4$ topology.}. 
  
We consider a generic action principle
describing the Einstein gravity coupled with some matter fields with a
lagrangean $L_m$
\begin{eqnarray}
S=\int d^4 x \sqrt{-g} ( \frac{1}{16 \pi G}R+L_m ),
\end{eqnarray}
which implies the Einstein equations
\begin{eqnarray}
R_{\mu \nu}-\frac{1}{2}R g_{\mu \nu}= 8 \pi G T_{\mu \nu},~~{\rm where~~~}
T_{\mu \nu}=-\frac{2}{\sqrt{-g}}\frac{\delta L_m}{\delta g^{\mu \nu}}~.
\end{eqnarray}
%
All known rotating regular solutions have been found
within the usual Lewis-Papapetrou ansatz  \cite{Wald-book} for a stationary,
axially symmetric spacetime with two Killing vector fields
$\partial/\partial  \varphi$ and $\partial/\partial t$.
A suitable parametrization of the metric line element 
in terms of the spherical coordinates $r,~\theta$ and $\varphi$,
used in most studies on this subject, is
\begin{equation}
ds^2=
  - f dt^2 +  \frac{m}{f} \left( d r^2+ r^2d\theta^2 \right)
           +  \frac{l}{f} r^2\sin^2\theta (d\varphi-\frac{\omega}{r} dt)^2
\ , \label{metric}
\end{equation}
where $f$, $m$, $l$ and $\omega$ are only functions of $r$ and $\theta$.

The asymptotic flatness imply that  the metric function are given at infinity by
$f=m=l=1,~~\omega=0.$
For solutions with a regular origin one has  
$\partial_r f=\partial_r m=\partial_r l=\omega=0 $ as $r\to 0$.
%
Also, all known examples of rotating solitons 
possess a parity symmetry with respect the $\theta=\pi/2$ plane.
Therefore, it is enough to solve the field equations for $0\leq \theta \leq \pi/2$;
the  derivatives
$\partial_\theta f$, $\partial_\theta l$, $\partial_\theta m$,
$\partial_\theta \omega$, have to vanish for both $\theta=0$ and $\theta=\pi/2$.

The mass $M$ and the angular momentum $J$ of the soliton solutions
can be read from the metric function expansion as $r \to \infty$ 
\begin{eqnarray}
\label{exp1}
 f \sim 1-\frac{2GM}{r}, ~~\omega \sim \frac{2GJ}{r^2},
\end{eqnarray}
 or equivalently from \cite{Wald-book}
 \begin{eqnarray}
\label{exp2}
M=-\int (2 T_t^t-T_{\mu}^{\mu}) \sqrt{-g} dr
d\theta d\varphi,~~~
 J=  \int T_{\varphi}^{t}\sqrt{-g} dr d\theta
d\varphi,
 \end{eqnarray}
the angular momentum being the charge associated with the Killing vector $\partial/\partial \varphi$.
 
 The existence and the properties of the rotating solitons depends on the 
 matter fields choice.
To find the features of  
these types of  configuration,
we have to solve extremely complicated partial differential 
equations, no closed form rotating soliton solution being known in the literature.

\section{Nonabelian rotating solitons}
We consider first the physically interesting case of 
a spontaneously broken gauge theory,
described by a matter lagrangean
 \begin{equation}
 \label{lm1}
-L_m=  Tr\{\frac{1}{2}F_{\mu \nu}F^{\mu \nu}\}+Tr\{\frac{1}{2}D_{\mu}\Phi D^{\mu}\Phi\}+
V(\Phi).
\end{equation}
with the Higgs field in the adjoint representation, 
$V(\Phi)=\frac{\lambda}{8} Tr(\Phi^2 - \eta^2)^2$ being the usual scalar potential.
The nonabelian field strength tensor is (here we restrict to a  SU(2)  gauge group)
\begin{eqnarray}
\nonumber
F_{\mu \nu}&=&\partial_{\mu}A_{\nu}-\partial_{\nu}A_{\mu}+i[A_{\mu},A_{\nu} ],
\end{eqnarray}
and the covariant derivative
$D_{\mu}=\partial_{\mu}+i[A_{\mu},\ ].$
Varying the action  with respect to
$A_{\mu}$ and $\Phi$ 
one finds the Yang-Mills-Higgs field equations 
\begin{eqnarray}
\label{YMeqs}
\frac{1}{\sqrt{-g}}D_{\mu}(\sqrt{-g} F^{\mu \nu}) &=&\frac{1}{4} i[\Phi,D^{\nu}\Phi],
\\
\label{Heqs}
\frac{1}{\sqrt{-g}}D_{\mu}(\sqrt{-g}D^{\mu} \Phi) +\lambda(\Phi^2-\eta^2)\Phi&=&0.
\end{eqnarray}
The variation of the lagrangian (\ref{lm1}) with respect
to the metric $g_{\mu \nu}$ yields the energy-momentum tensor
which enters the Einstein equations
\begin{eqnarray}
\label{tensor}
T_{\mu\nu} = 2Tr\{F_{\mu \alpha} F_{\nu \beta} g^{\alpha \beta} 
-\frac{1}{4}g_{\mu \nu}F_{\alpha \beta}F^{\alpha \beta} \}
+Tr\{\frac{1}{2}D_{\mu}\Phi D_{\nu}\Phi 
-\frac{1}{4}g_{\mu \nu}(D_{\alpha}\Phi D^{\alpha}\Phi\}+
V(\Phi)g_{\mu\nu}.
\end{eqnarray}
We mention also the expression for the electric and magnetic charges
derived by using the 't Hooft field strength tensor (with  $\hat{\Phi}=\Phi/|\Phi|$)
\begin{eqnarray}
\label{e-charge}
Q_e=\frac{1}{4\pi}\oint_{\infty}dS_{\mu}Tr\{ \hat{\Phi}F_{\mu t} \},
~~
Q_m=\frac{1}{4\pi}\oint_{\infty}dS_{\mu}\frac{1}{2}\epsilon_{\mu \nu \alpha}
Tr\{\hat{\Phi}F_{\nu \alpha}\}.
\end{eqnarray}

In the purely magnetic case (no electric potential
$A_t=0$), the Einstein-Yang-Mills-Higgs (EYMH) field equations admits a variety of
solutions, the gravitating version of the spherically symmetric
't Hooft-Polyakov monopoles 
being the best known example.
Of interest here are also the composite axially symmetric solutions containing magnetic
charges of both signs. The simplest case consists in two opposite charges
located on the $z-$axis and forming a monopole-antimonopole pair, 
$i.e.$ a magnetic dipole \cite{Kleihaus:2000hx}.

The issue of rotating solutions in EYMH model was first addressed
within a perturbative approach in the nongravitating limit.
However, in the absence of gravity, 
it has been shown that 't Hooft Polyakov monopoles and Julia-Zee dyons do not
admit  slowly rotating excitations \cite{Heusler:1998ec}.

Further progress in this area was made possible by the discovery that
the angular momentum $J$ in (\ref{exp2}) admits a 
simple expression as a surface integral 
in terms of YM fields only \cite{VanderBij:2001nm}.
One starts by noticing that, at the level of the matter ansatz, a rotation around the $z-$axis
can be compensated by a gauge rotation
\begin{eqnarray} \label{Psi}
{\mathcal{L}}_\varphi A=D\Psi,
\end{eqnarray}
and therefore
$F_{\mu \varphi} =  D_{\mu}W,
~~
D_{\varphi}\Phi= i[W,\Phi] ,$
where $W=A_{\varphi}-\Psi$.

Therefore one may write the following expression for the $T_{\varphi}^{t}$-component of the
energy-momentum tensor associated with rotation
\begin{eqnarray}
T_{\varphi}^{t}&=&2Tr\Big\{ (D_{r}W)F^{rt}
+(D_{\theta}W)F^{\theta t}
+\frac{i}{4}[W,\Phi] D^{t} \Phi\Big\}
\nonumber\\
&=&2Tr\Big\{\frac{1}{\sqrt{-g}}D_{r}(WF^{rt}\sqrt{-g})
+\frac{1}{\sqrt{-g}}D_{\theta}(WF^{\theta t}\sqrt{-g})
\nonumber\\
&^{}&-W\Big(
\frac{1}{\sqrt{-g}}D_{r}(\sqrt{-g}F^{rt})
+\frac{1}{\sqrt{-g}}D_{\theta}(\sqrt{-g}F^{\theta t})\Big)
+\frac{i}{4}[W,\Phi] D^{t} \Phi\Big\}.
\end{eqnarray}
As a consequence of the YM equations (\ref{YMeqs})
and making use of the fact that the trace
of a commutator vanishes we obtain
\begin{eqnarray}
\label{T34}
\nonumber
T_{\varphi}^{t}=2Tr\Big\{\frac{1}{\sqrt{-g}}\partial_{\mu}(WF^{\mu t}\sqrt{-g})\Big\}.
\end{eqnarray}
Thus, for globally regular solutions, 
the total angular momentum can be expressed as an integral
over the two-sphere at spacelike infinity \cite{VanderBij:2001nm, vanderBij:2002sq}
\begin{eqnarray}
\label{totalJ}
J =\oint_{\infty}2Tr\{WF^{\mu t} \} dS_{\mu}
=2 \pi \lim_{r \rightarrow \infty}  \int_0^{\pi} d \theta \sin \theta^{~}
 r^2[W^{(r)}F^{rt(r)}+W^{(\theta)}F^{rt(\theta)}+W^{(\varphi)}F^{rt(\varphi)}].
\end{eqnarray}
This generic relation is evaluated
for a specific axially symmetric ansatz, within a set of boundary conditions consistent
with finite energy and regularity assumptions.
 
As proven in  \cite{VanderBij:2001nm, vanderBij:2002sq}, the angular momentum of a gravitating  
monopole-antimonopole
solution is nonzero and equals the electric charge $Q_e$.
In fact, it appears to exist a very general (-and still poor understood) connection
between the angular momentum and the topological charge of an 
axially symmetric, ellectrically charged
solution in EYMH theory.
The total angular momentum of any solution with a nonvanishing global magnetic charge
is zero, although the configurations
rotates locally, $g_{t \varphi} \neq 0$. 
The solutions without a global topological charge (which presents, however, a
nonvanishing magnetic charge density) have a nonzero angular momentum proportional to the 
electric charge, 
\begin{equation}
 J=nQ_e,
\end{equation} 
where $n$ is an integer - the winding number of solutions.

The only numerical solutions exhibited so far in literature correspond to the charged monopole
\cite{Kleihaus:2005fs}  and
monopole-antimonopole  \cite{Paturyan:2004ps} cases. 
They were found within 
a suitable parametrization of the
axially symmetric ansatz 
derived by Rebbi and Rossi
\cite{RR}, with a SU(2) gauge connection
\begin{equation}
\label{matter-ansatz}
\nonumber
 A_\mu dx^\mu =\vec A \cdot  d \vec r+A_t dt= \frac{1}{2er}
\left[ \tau _\phi
 \left( H_1 dr + \left(1-H_2\right) r d\theta \right)
 -\left( \tau_r H_3 + \tau_\theta \left(1-H_4\right) \right)
  r \sin \theta d\varphi
+\left( \tau_r H_5 + \tau_\theta H_6 \right) dt \right] \ ,
\end{equation}
and a  Higgs field of the form
\begin{equation}
\nonumber
\Phi= \left(\phi_1 \tau_r+\phi_2 \tau_\theta\right)
\ . \end{equation}
The SU(2) matrices $(\tau_r,\tau_\theta,\tau_\phi)$
are defined in terms of the Pauli matrices
$\vec \tau = ( \tau_x, \tau_y, \tau_z) $ by
$\tau_r = \vec \tau \cdot
(\sin \theta \cos n\varphi, \sin \theta \sin  n\varphi, \cos \theta)$,
$\tau_\theta = \vec \tau \cdot
(\cos \theta \cos n \varphi, \cos \theta \sin  n\varphi, -\sin \theta)$,
$\tau_\phi = \vec \tau \cdot (-\sin  n\varphi, \cos n \varphi,0)$.

%
%
%
%
\newpage
\setlength{\unitlength}{1cm}
\begin{picture}(4,8)
\centering
\put(1,1){\epsfig{file=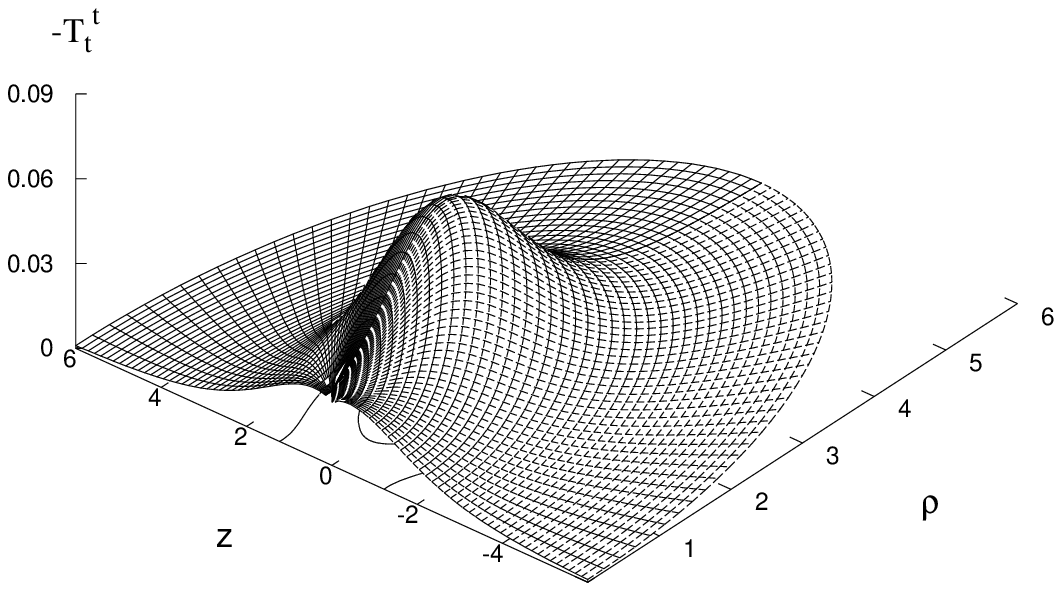,width=14cm}}
\end{picture}
\begin{picture}(20,8)
\centering
\put(1.5,1){\epsfig{file=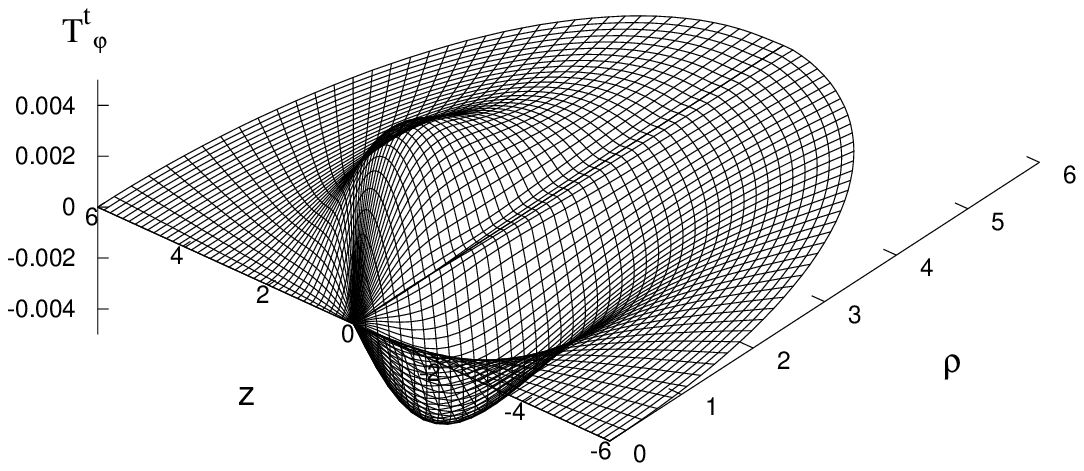,width=14cm}}
\end{picture}
{\small {\bf Figure 1.} The mass-energy density $\epsilon=-T_t^t$ and the angular momentum 
density $T_\varphi^t$  of a typical rotating dyon solution in EYMH theory
are plotted 
as a function of the  coordinates $\rho =   r \sin \theta$ and
$z =  r \cos \theta$.}
\\
\\
The six gauge field functions $H_i$ and the two Higgs field function
$\phi_i$ depend only on the coordinates $r$ and $\theta$.

The boundary conditions they satisfy depend on the presence 
or not of a global magnetic charge and are
 essentially fixed by the asymptotic behaviour of the Higgs field
$|\Phi| \to \eta$   
\begin{equation}
\label{phi-rinfty}
\lim_{r\to\infty}\phi_1=\eta \cos m\theta,~~~
\lim_{r\to\infty}\phi_2=\eta \sin m\theta\,,
\end{equation}
with $m=0,1,..,$ plus regularity requirements.  
The solutions with even $m$ have a nonzero magnetic charge and $J=0$; 
odd $m$ solutions have a different picture,
with $Q_m=0$ and a nonzero angular momentum.

  Solving the field equations reveals a complicated branch structure.
For both dyons and electrically charged monopole-antimonopoles, 
a branch of rotating solutions emerges from the flat spacetime
configurations.
\newpage
\setlength{\unitlength}{1cm}
\begin{picture}(4,8)
\centering
\put(1,1){\epsfig{file=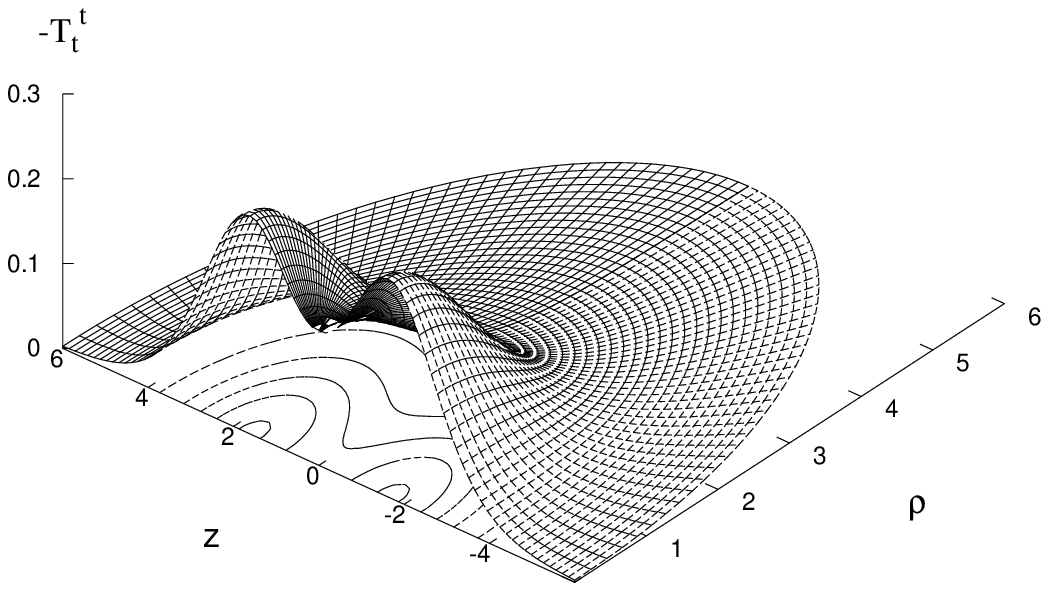,width=13cm}}
\end{picture}
\begin{picture}(20,8.5)
\centering
\put(1.7,1){\epsfig{file=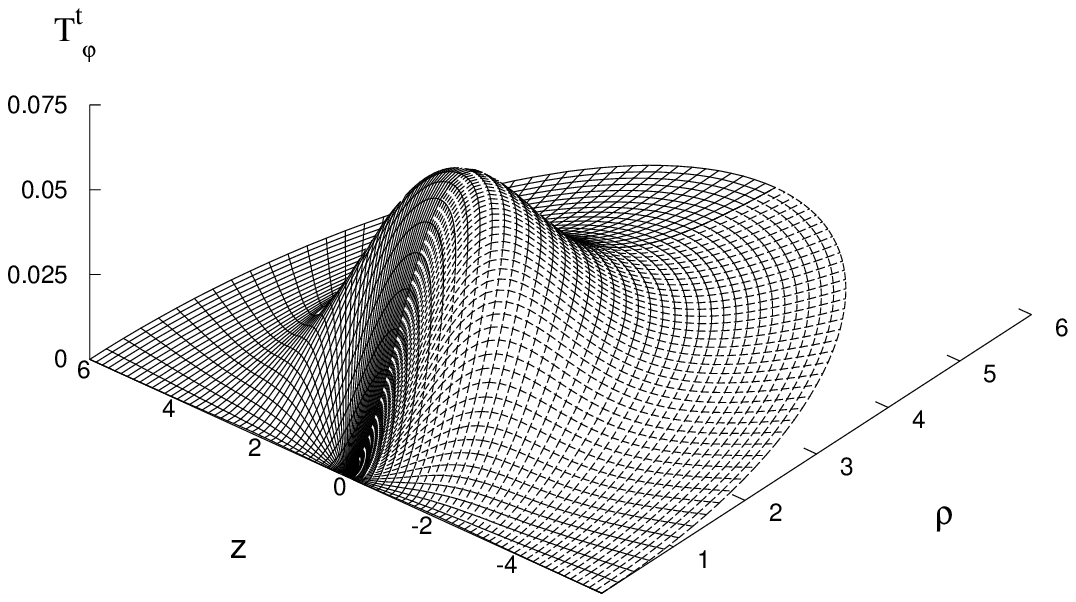,width=13cm}}
\end{picture}
{\small {\bf Figure 2.} The  same as Figure 1 for a typical rotating monopole-antimonopole solution.}
\\
\\
Apart from this fundamental branch, there are also excited solutions 
related to axially symmetric solutions of
EYM theory
(for more details, see \cite{Kleihaus:2005fs,Paturyan:2004ps}).

In Figures 1, 2 we show the energy density $\epsilon=-T_t^t$ and the
angular momentum density $T^t_{\varphi}$ for typical gravitating 
dyon and monopole-antimonopole solutions
as function of the coordinates $z=r \cos\theta$ and $\rho=r
\sin\theta$.
As seen from these  Figures,
the distributions of the dyon mass-energy density
shows a  peak along the $\rho$-axis
and decreases monotonically along the $z$-axis.
Equal density contours reveal a torus-like shape
of the solutions.
The picture is different for the angular momentum density which vanishes on the
 $\rho$-axis and changes the sign as $z \to -z$.
Thus, although it will rotate locally, the total angular
momentum of the dyon solutions is zero (although 
 $g_{\varphi t} \neq 0$), and the spacetime consists in two
regions rotating in opposite directions. 

The energy density
of a monopole-antimonopole configuration possesses two maxima symmetrically located on the $z-$axis
 and a saddle point at the origin,
 describing a composite configuration. 
 A different picture is found for the
angular momentum density. As seen in Figure 2, the magnetic dipole 
system rotates as a
single object and the $T_\varphi^t$-component of the energy momentum
tensor associated with rotation presents a maximum in the $z=0$ plane and
no local extrema at the locations of the monopole and the antimonopole.

However, it is interesting that no rotating solutions 
are found in the limit of vanishing Higgs field.
Spinning generalizations of the BK solitons in EYM theory, although
predicted perturbatively \cite{Brodbeck:1997ek},
appears do not exist within a nonperturbative approach \cite{vanderBij:2002sq}.
For such configurations, the $A_{t}$-components of the gauge 
field act like an isotriplet Higgs field 
with negative metric, and by themselves cause the magnetic components of the 
nonabelian potential
to oscillate rather than decrease exponentially as $r \to \infty$,
which would give an infinite mass.
Therefore we are forced to take  
$\lim_{r \to \infty}A_{t}=0$. However, this implies a vanishing electric field,
$i.e.$ a zero Poynting vector and a static configuration $g_{\varphi t}=0$
\footnote{ Rotating black hole solutions in EYM model are known to exist \cite{Kleihaus:2002ee}.
However, these configurations are found within a set of boundary conditions with a vanishing electric
electric potential at infinity and are sustained by the existence of an event horizon.
}.

Here we should remark that another example of a four dimensional
rotating soliton was found recently in a closed related
model presenting a U(1) gauge field and a dilaton interacting in a nontrivial way with 
the Yang-Mills and Higgs fields \cite{Brihaye:2005pz}.
This model originates in Kaluza-Klein reduction of the five dimensional EYM theory.
The $d=4$ rotating solitons are found by reducing along the $x^5$-direction the 
boosted five
dimensional static axisymmetric nonabelian vortices.
As expected, the resulting rotating configurations share many features with
the EYMH dyons and monopole-antimonopole solutions discussed above.

\section{Spinning U(1) gauged Skyrmions}
To the best of our knowledge, 
the only example of rotating solution residing
in the one-soliton sector of the theory which has a
topologically stable limit was found in U(1) gauged Skyrme theory.   

The Skyrme model has been proposed a long time ago \cite{Skyrme:1961vq}
as an effective theory for nucleons in the large $N$ limit 
of QCD at low energy~\cite{Witten:1983tx},
the baryon number being identified with 
the topological charge.
The classical as well as the quantum properties are 
in relatively good agreeement with the observed features of small nuclei.
The $U(1)$ gauged Skyrme model
 was originally proposed by Callan and Witten to study the
decay of the nucleons in the vecinity of a monopole \cite{Callan:1983nx}.

In terms of the $O(4)$ sigma model field $\f^a=(\f^{\al},\f^A)$, $\al=1,2$;
$A=3,4$, satisfying the constraint $|\f^{\al}|^2+|\f^A|^2=1$, the Lagrangean
of the Maxwell gauged Skyrme model is 
(up to an overall factor which we set equal to one)
\be
\label{lagu1}
-L_m=\frac12 F_{\mu\nu}F^{\mu \nu}+\frac12 D_{\mu}\f^a  D^{\mu}\f^a
+\frac{\ka^2}{8}\,|D_{[\mu}\f^a\,D_{\nu]}\f^b|^2
\ee
in terms of the Maxwell field strength $F_{\mu\nu}$, and the covariant
derivatives defined by the gauging prescription \cite{Piette:1997ny}
\be
\label{u1pres}
D_{\mu}\f^{\al}=\pa_{\mu}\f^{\al}+A_{\mu}\,(\vep\f)^{\al}\quad,\quad
D_{\mu}\f^A=\pa_{\mu}\f^A\,.
\ee
The energy-momentum tensor which follows from (\ref{lagu1}) is
\bea
T_{\mu\nu}&=&\Bigg\{2\left(F_{\mu\la}\,F_{\mu}^{\la}
-\frac14\,g_{\mu\nu}\,F_{\tau\la}\,F^{\tau\la}\right)+
\left(D_{\mu}\f^a\,D_{\nu}\f^a
-\frac12\,g_{\mu\nu}\,D_{\la}\f^a\,D^{\la}\f^a\right)\nonumber\\
&+&2\cdot\frac{\ka^2}{4}\left[\left(D_{[\mu}\f^a\,D_{\la]}\f^b\right)
\left(D_{[\nu}\f^a\,D^{\la]}\f_b\right)-\frac14\,g_{\mu\nu}
\left(D_{[\tau}\f^a\,D_{\la]}\f^a\right)
\left(D^{[\tau}\f_a\,D^{\la]}\f_b\right)\right]\Bigg\}\,.\label{stress-mn}
\eea
The rotating gauged Skyrmions are found within the following matter ansatz
\footnote{A nongravitating axially 
symmetric, spinning soliton of the ungauged Skyrme model, 
has been recently constructed in \cite{Battye:2005nx}. However, this is a
$Q$-ball type of solution featuring time-dependent fields.}
\begin{eqnarray}
&A_{\mu}dx^\mu=(a(r,\theta)-n)d\varphi+b(r,\theta) dt,
\\
\nonumber
&\f^{\al}=\sin F(r,\theta)\sin G(r,\theta)\ n^{\al},
~~~
\f^3=\sin F(r,\theta)\cos G(r,\theta) ,
~~~\f^4=\cos F(r,\theta),
\end{eqnarray}
\newpage
\setlength{\unitlength}{1cm}
\begin{picture}(4,8)
\centering
\put(1,1){\epsfig{file=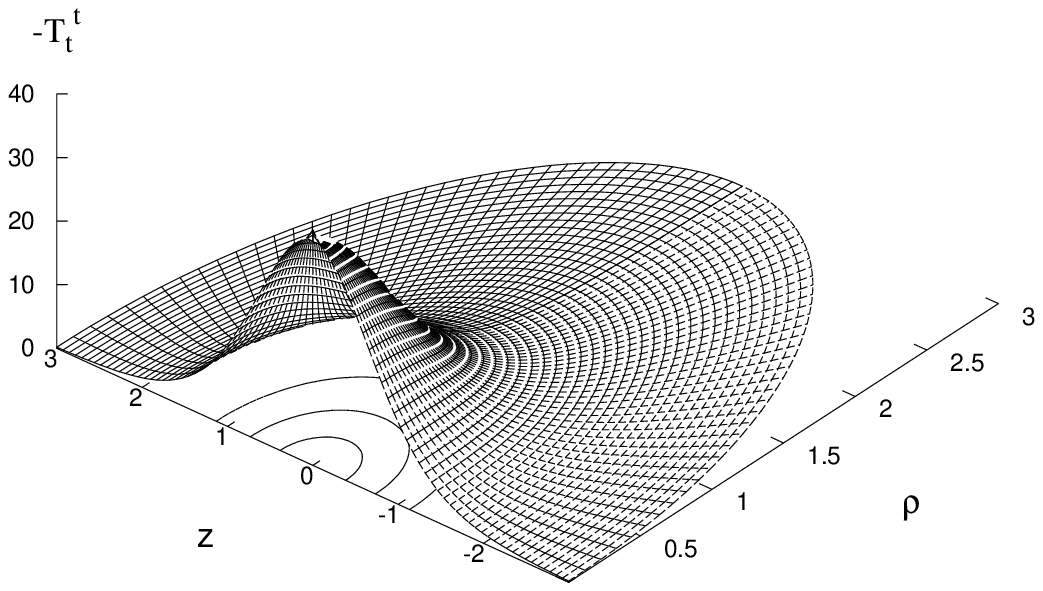,width=14cm}}
\end{picture}
\begin{picture}(20,8.5)
\centering
\put(1.7,1){\epsfig{file=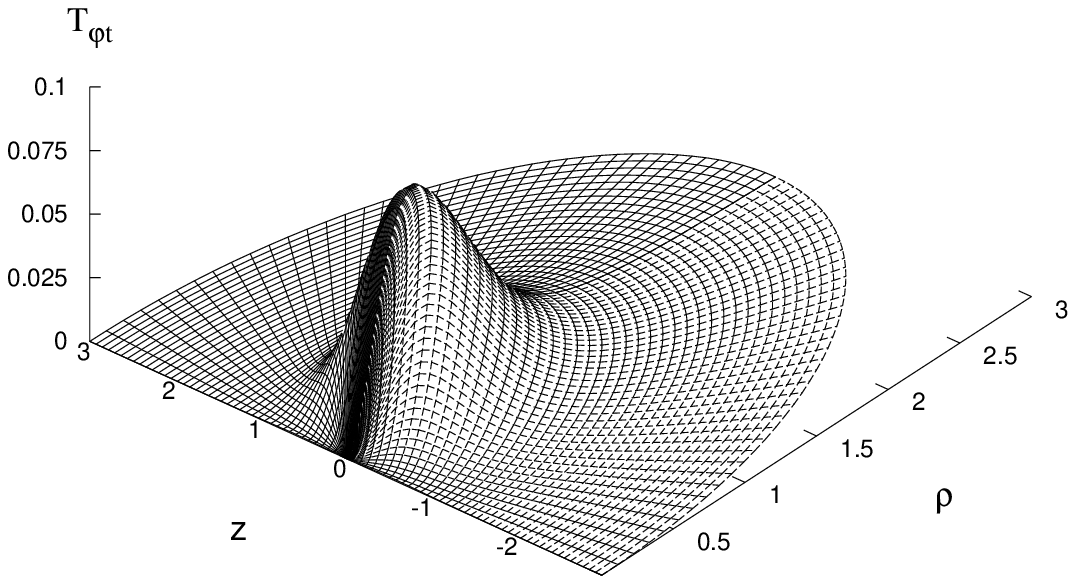,width=14cm}}
\end{picture}
{\small {\bf Figure 3.} The mass-energy density $\epsilon=-T_t^t$ and the angular momentum 
density $T_\varphi^t$  of a typical rotating gauged Skyrmeon solution.}
\\
\\
$a$ and $b$ corresponding to the magnetic and electric components of the 
abelian potential, with $n$ a positive integer - the Baryon number.

By using the field equations, the volume integral of  
$T_{\varphi}^t$ can be converted into a surface integral at infinity
in terms of Maxwell potentials
\begin{eqnarray}
J=4\pi \lim_{r \rightarrow \infty} \int_{0}^{\pi}d\theta \sin \theta~r^2 a~b_r.
\end{eqnarray}
The field equations imply the  asymptotic behaviour of
the  electric potential, 
$b\sim V-Q/(2r)+O(1/r^2), $
the parameter $Q$ corresponding to the electric charge of the solutions.
Therefore the following relation holds
\begin{eqnarray}
\label{JQ}
J= n Q.
\end{eqnarray} 
Note that the solutions discussed here 
possess also a magnetic dipole moment 
which can be read from the asymptotics of the $U(1)$ 
magnetic potential, $A_{\varphi}\sim \mu \sin\theta/r^2$, while the magnetic charge is zero.

This system has been discussed in the flat spacetime limit in Ref. \cite{Radu:2005jp}, 
where the basic properties
of the configurations are expossed.
Gravitating generalizations of these solutions can easily be constructed \cite{PRT}.
For a given Baryon number, 
the solutions depend on two continuos parameters, the values $V$
of the electric potential at infinity and the 
Skyrme coupling constant $\kappa$.
The solutions with $V=0$ have $b=0$ and correspond 
to static dipoles.
A nonvanishing $V$ leads to rotating regular configurations,
with nontrivial functions
$f,~l,~m$, $\omega$ and $F,~G,~a,~b$.

In Figure 3 we plot the energy density   and  
the angular momentum density 
of a typical gravitating  $n=1$ Skyrmion solution as a function of the
coordinates $\rho, z$.
We notice that the energy density  does not
exhibit any distinctly localised individual components, a surface
of constant energy density being
topologically a sphere. 
Also, the electrically charged $U(1)$ gauged Skyrmion rotates as a
single object and the $T_{\varphi}^t$--component of the energy-momentum
tensor associated with rotation presents a maximum in the $z=0$ plane and
no local extrema.
%
%
\section{Rotating boson stars}
The first example of rotating soliton in general relativity
was found in  a theory containing a  
complex scalar field with an harmonic time dependence.

Spherically symmetric, gravitationally bound 
states of scalar field were first obtained by Kaup \cite{Kaup}
and Ruffini and Bonazzola \cite{Ruffini:1969qy}.
These boson stars are macroscopic quantum states and are only prevented 
from collapsing
gravitationally by the Heisenberg uncertainty principle
(see \cite{Schunck:2003kk} for a recent review of this type of soliton solutions).
 
The lagrangian density of a complex self-gravitating scalar field $\Phi$ reads
\begin{eqnarray}
\label{boson}
-L_{m}=\sqrt{-g }\left( g^{i j}\Phi,_{i}^{\ast} \Phi_{j},
+ V(\Phi)\right),
\end{eqnarray}
where the asterisc denotes complex conjugate.
Here we consider only the  case $V(\Phi)=\mu^2 \Phi^{\ast}\Phi$,
where $\mu$ is the scalar field mass 
\footnote{ A potential on the form $V(\Phi)=\lambda (|\Phi|^6-a |\Phi|^4)+b|\Phi|^2)$
leads to nontopological soliton solutions (Q-balls) which exist even in the absence of
gravity. Rotating solutions of this model are discussed in \cite{Volkov:2002aj}.}.

The Lagrangian density (\ref{boson}) is invariant under a global phase rotation 
$\Phi \to \Phi e^{-i\alpha}$; 
that implies the existence of a conserved current 
\begin{eqnarray}
\label{J}
J^{k}=i g^{kl}
\left (\Phi^{\ast}_{,l}\Phi - \Phi_{,l}\Phi^{\ast} \right),
\end{eqnarray}
and an associated conserved charge, namely, the number of scalar particles
\begin{eqnarray}
N=\int d^{3} x \sqrt{-g } J^t.
\end{eqnarray}
The energy momentum tensor is given by
\begin{eqnarray}
\label{Tij}
T_{ij}=\Phi^{\ast}_{,i}\Phi_{,j}+\Phi^{\ast}_{,j}\Phi_{,i}
-g_{ij}(g^{km}\Phi^{\ast}_{,k}\Phi_{,m}+\mu^2 |\Phi|^2).
\end{eqnarray}
Rotating boson star solutions are found for a scalar field ansatz
\begin{eqnarray}
\Phi=\phi(r,\theta) e^{i(m\varphi-\omega t)}.
\end{eqnarray}
Single-valuedness of the scalar field  requires
$\Phi(\varphi)= \Phi(\varphi+2\pi)$.
Thus the constant $m$ must be an integer $m=0,\pm 1,\dots$.

Rotating boson stars solutions have been considered 
by various authors \cite{boson}, \cite{Kleihaus:2005me}, \cite{Schunck},
these being the best understood examples of spinning configurations.
One can easily see that
the angular momentum of these solutions is  quantized
\begin{eqnarray}
J=m N.
\end{eqnarray} 
The results in the literature indicate that rotating boson stars exist only 
for a limited frequency range $\omega_{min}<\omega<\omega_{max}$.
The mass $M$ and particle number $N$ tends to zero when the maximal frequency
is approached.
For each rotational quantum number $m$, there are even and odd parity
solutions, although only even parity rotating solutions have been exhibited 
in the literature.

A boson star rotates as a single object, without being
possible to distinct localised individual components.
The typical profiles of the energy density and angular momentum density 
have a similar shape with those presented in Figure 3.
\section{Further remarks}
Recently, some progress was achieved in understanding the nature
of rotating soliton solutions in general relativity.
Several different models have been examined in literature, with a number of surprising results.

A general feature of all known rotating solitons is that the angular momentum
is fixed by some conserved charge of the matter fields 
(the electric charge in the presence of a gauge field or the particle number
for a complex scalar field with an harmonic time dependence).

Concerning the case of a spontaneously broken gauge theory, 
a deep connection appears to exist between the global magnetic charge and angular momentum.
The rotating solitons of this theory 
have a vanishing magnetic charge and are unstable.
Also, we expect the EYMH theory to present a
whole sequence of rotating solutions
generalizing for a nonzero electric potential the magnetic
chains and rings found in \cite{Kleihaus:2004is}.
These configurations will satisfy the generic relation 
between the nonabelian charges and angular momentum.

We should also remark that all known rotating solitons in general relativity
are curved space generalizations
of flat space rotating configurations \footnote{The boson star solutions
are not a genuine counterexample given the 
existence of flat space Q-ball solutions with many similar properties.}.
No  example of rotating soliton sustained by gravity is known in the literature.
In particular,
 although
predicted perturbatively, no rotating
generalisations of the BK solution seem to exist. 
However, recently it has been realized that, apart from BK solution, the EYM equations
admit a general set of static axially symmetric configurations \cite{Ibadov:2004rt}.
Although we expect that no rotating generalizations will be found in this case also,
this issue may deserve a careful study.
 
Considering the case of a double Higgs field, the authors of \cite{Volkov:2003ew} 
concluded that the well known SU(2)
sphalerons do not admit spinning generalizations within the stationary, axially
symmetric ansatz. However, similar to the case of a Higgs field in the adjoint representation,
this argument does not exclude the existence of more complex, composed configurations
which  possible may rotate.

It would be interesting to look for rotating solitons in other models admiting
particle-like solutions.
Particularly interesting are various supersymmetric theories, which may lead to new 
qualitative features.
Also, very little is known on the question of rotating solitons with higher gauge groups.

The question of rotating solutions may still hold further surprises.
For example, all know solutions have been found  within the  Lewis-Papapetrou ansatz 
(\ref{metric}).
However, the inclusion of nonabelian matter fields
leaves open the possibility of existence of more general rotating solutions,
exciting other extradiagonal metric components as well \cite{Heusler:1996ft}.

We close by remarking that the issue of rotating solitons crucially depends  
on asymptotic structure of spacetime.
For example, the Melvin magnetic geon \cite{Melvin:1963qx} 
is known to possess rotating generalizations \cite{Garfinkle:1994wk}.
Also, the EYM-SU(2) system presents rotating soliton solutions \cite{Radu:2002rv}
for anti-de Sitter asymptotics.
\\ 
\\
{\bf Acknowledgement}
\\
This work was carried out in the framework of Enterprise--Ireland
Basic Science Research Project SC/2003/390.
\newpage




\begin{thebibliography}{99}
\bibitem{Lichner}
A. Lichnerowicz,
 {\it Th\'eories Relativiste de la
  Gravitation et de l'\'Electromagn\'etisme},
Masson, Paris, 1955.
\bibitem{mon}
G.'t~Hooft,     Nucl. Phys. B 
 {\bf 79}, 276 (1974);
\\ 
A.M.~Polyakov,  
 JETP Lett.   {\bf 20}, 430 (1974).
\bibitem{dyon}
B.~Julia, A.~Zee,    Phys. Rev. D {\bf 11}, 2227 (1975).
\bibitem{sphal}
F.R.~Klinkhamer and N.S.~Manton,  
 Phys. Rev. D
{\bf  30}, 2212 (1984).
\bibitem{Skyrme:1961vq}
  T.~H.~R.~Skyrme,
  Proc.\ Roy.\ Soc.\ Lond.\ A {\bf 260} (1961) 127;
  \\
  T.~H.~R.~Skyrme,
  Nucl.\ Phys.\  {\bf 31} (1962) 556.
\bibitem{Bartnik:1988am}
  R.~Bartnik and J.~Mckinnon,
  Phys.\ Rev.\ Lett.\  {\bf 61} (1988) 141.
\bibitem{Kaup}
D. J. Kaup,
Phys.\ Rev.\  {\bf 172} (1968) 1332.
\bibitem{Ruffini:1969qy}
R.~Ruffini and S.~Bonazzola,
Phys.\ Rev.\  {\bf 187} (1969) 1767.

\bibitem{Wald-book}
R. Wald, {\it General Relativity}, University of Chicago Press, Chicago, 1984.
 
\bibitem{Kleihaus:2000hx}
  B.~Kleihaus and J.~Kunz,
  Phys.\ Rev.\ Lett.\  {\bf 85} (2000) 2430
  [arXiv:hep-th/0006148].

\bibitem{Heusler:1998ec}
  M.~Heusler, N.~Straumann and M.~S.~Volkov,
  Phys.\ Rev.\ D {\bf 58} (1998) 105021
  [arXiv:gr-qc/9805061].
\bibitem{VanderBij:2001nm}
  J.~J.~Van der Bij and E.~Radu,
  Int.\ J.\ Mod.\ Phys.\ A {\bf 17} (2002) 1477
  [arXiv:gr-qc/0111046].
\bibitem{vanderBij:2002sq}
  J.~J.~van der Bij and E.~Radu,
  Int.\ J.\ Mod.\ Phys.\ A {\bf 18} (2003) 2379
  [arXiv:hep-th/0210185].

\bibitem{Kleihaus:2005fs}
  B.~Kleihaus, J.~Kunz and U.~Neemann,
  Phys.\ Lett.\ B {\bf 623} (2005) 171
  [arXiv:gr-qc/0507047].
\bibitem{Paturyan:2004ps}
  V.~Paturyan, E.~Radu and D.~H.~Tchrakian,
  Phys.\ Lett.\ B {\bf 609} (2005) 360
  [arXiv:hep-th/0412011].
\bibitem{RR}
C. Rebbi, P. Rossi, Phys.Rev.{\bf D22} (1980) 2010.  
\bibitem{Brodbeck:1997ek}
  O.~Brodbeck, M.~Heusler, N.~Straumann and M.~S.~Volkov,
  Phys.\ Rev.\ Lett.\  {\bf 79} (1997) 4310
  [arXiv:gr-qc/9707057].
\bibitem{Kleihaus:2002ee}
  B.~Kleihaus, J.~Kunz and F.~Navarro-Lerida,
  Phys.\ Rev.\ D {\bf 66} (2002) 104001
  [arXiv:gr-qc/0207042].
\bibitem{Brihaye:2005pz}
  Y.~Brihaye, B.~Hartmann and E.~Radu,
  Phys.\ Rev.\ D {\bf 71} (2005) 085002
  [arXiv:hep-th/0502131].
\bibitem{Witten:1983tx}
  E.~Witten,
  Nucl.\ Phys.\ B {\bf 223} (1983) 433;
\\
  E.~Witten,
  Nucl.\ Phys.\ B {\bf 223} (1983) 422;
\\  
  G.~S.~Adkins, C.~R.~Nappi and E.~Witten,
  Nucl.\ Phys.\ B {\bf 228} (1983) 552.
\bibitem{Callan:1983nx}
  C.~G.~Callan and E.~Witten,
  Nucl.\ Phys.\ B {\bf 239} (1984) 161.
\bibitem{Piette:1997ny}
  B.~M.~A.~Piette and D.~H.~Tchrakian,
  Phys.\ Rev.\ D {\bf 62} (2000) 025020
  [arXiv:hep-th/9709189].
\bibitem{Battye:2005nx}
  R.~A.~Battye, S.~Krusch and P.~M.~Sutcliffe,
  Phys.\ Lett.\ B {\bf 626} (2005) 120
  [arXiv:hep-th/0507279].
\bibitem{Radu:2005jp}
  E.~Radu and D.~H.~Tchrakian,
  arXiv:hep-th/0509014.
\bibitem{PRT}
V. Paturyan, E. Radu, D. H. Tchrakian, {\it to appear}. 
\bibitem{Schunck:2003kk}
  F.~E.~Schunck and E.~W.~Mielke,
  Class.\ Quant.\ Grav.\  {\bf 20} (2003) R301.
\bibitem{Volkov:2002aj}
  M.~S.~Volkov and E.~Wohnert,
  Phys.\ Rev.\ D {\bf 66} (2002) 085003
  [arXiv:hep-th/0205157].

\bibitem{boson}
S. Yoshida, Y. Eriguchi, Phys.Rev.{\bf D56} (1997) 762.  
\bibitem{Kleihaus:2005me}
  B.~Kleihaus, J.~Kunz and M.~List,
  Phys.\ Rev.\ D {\bf 72} (2005) 064002
  [arXiv:gr-qc/0505143].
\bibitem{Schunck}
F.~E.~Schunck, E.~W.~Mielke
Phys.\ Lett.\ A {\bf 249} (1998) 389;
\\
F.~E.~Schunck and E. W.~Mielke,
Gen. Rel. Grav. {\bf 31} (1999), 787.
\bibitem{Kleihaus:2004is}
  B.~Kleihaus, J.~Kunz and Y.~Shnir,
  Phys.\ Rev.\ D {\bf 70} (2004) 065010
  [arXiv:hep-th/0405169].
\bibitem{Ibadov:2004rt}
  R.~Ibadov, B.~Kleihaus, J.~Kunz and Y.~Shnir,
  Phys.\ Lett.\ B {\bf 609} (2005) 150
  [arXiv:gr-qc/0410091].
\bibitem{Volkov:2003ew}
  M.~S.~Volkov and E.~Wohnert,
  Phys.\ Rev.\ D {\bf 67} (2003) 105006
  [arXiv:hep-th/0302032].
\bibitem{Heusler:1996ft}
  M.~Heusler,
  Helv.\ Phys.\ Acta {\bf 69} (1996) 501
  [arXiv:gr-qc/9610019].
\bibitem{Melvin:1963qx}
  M.~A.~Melvin,
  Phys.\ Lett.\  {\bf 8} (1964) 65.
\bibitem{Garfinkle:1994wk}
  D.~Garfinkle and M.~A.~Melvin,
  Phys.\ Rev.\ D {\bf 50} (1994) 3859
  [arXiv:gr-qc/9404016].
\bibitem{Radu:2002rv}
  E.~Radu,
  Phys.\ Lett.\ B {\bf 548} (2002) 224
  [arXiv:gr-qc/0210074].

  
\end{thebibliography}
\end{document}